\documentclass[11pt]{article}
\usepackage{moriond,epsfig}

\def\Journal#1#2#3#4{{#1} {\bf #2}, #3 (#4)}

\def\JPG{{\em J. Phys.} G}

\def\NPA{{\em Nucl. Phys.} A}

\def\PLB{{\em Phys. Lett.}  B}

\def\PRD{{\em Phys. Rev.} D}
\def\PRC{{\em Phys. Rev.} C}

\def\be{\begin{equation}}
\def\ee{\end{equation}}
\def\bea{\begin{eqnarray}}
\def\eea{\end{eqnarray}}

\begin{document}
\vspace*{4cm}
\title{Collective flow in (anti)proton-proton collision at
Tevatron and LHC}

\author{ Tanguy$\,$Pierog$^{(1)}$, S.$\,$Porteboeuf$^{(2)}$, Iu.$\,$Karpenko$^{(3,4)}$, K.$\,$Werner$^{(4)}$}

\address{$^{(1)}$ Karlsruhe Institut of Technology, Institut f\"ur Kernphysik,
Postfach 3640, 76021 Karlsruhe Germany\\
$^{(2)}$ LLR, Ecole Polytechnique, Palaiseau, France\\
$^{(3)}$ Bogolyubov Institute for Theoretical Physics, Kiev 143, 03680, Ukraine\\
$^{(4)}$ SUBATECH, University of Nantes -- IN2P3/CNRS-- EMN, Nantes, France}

\maketitle\abstracts{
Collective flow as a consequence of hydrodynamical evolution in heavy ion 
collisions is intensively studied by theorists and experimentalists to
understand the behavior of hot quark matter. Due to their large mass, 
heavy ions suffer collective effects even at low (SPS) or
intermediate energies (RHIC). In case of light systems such as 
(anti)proton-proton interactions, collective effects was not expected. 
Within a global model such as EPOS, where light and heavy systems are
treated using the same physics, it appears that  Tevatron data are better 
described if a flow is introduced. Then the extrapolation to LHC can 
easily be done and we can compare to first data from ATLAS experiment.}

\section{Introduction}

There seems to be little doubt that heavy ion collisions at RHIC energies
produce matter which expands as an almost ideal fluid~\cite{intro1,intro4}.
This observation is mainly based on the studies of azimuthal anisotropies,
which can be explained on the basis of ideal hydrodynamics~\cite{hydro1}.
A big success of this approach was the correct description of the
so-called mass splitting, which refers to quite different transverse
momentum dependencies of the asymmetries for the different hadrons,
depending on their masses. 

\begin{figure}[hptb]
\begin{center}
\includegraphics[width=0.38\textwidth,angle=-90]{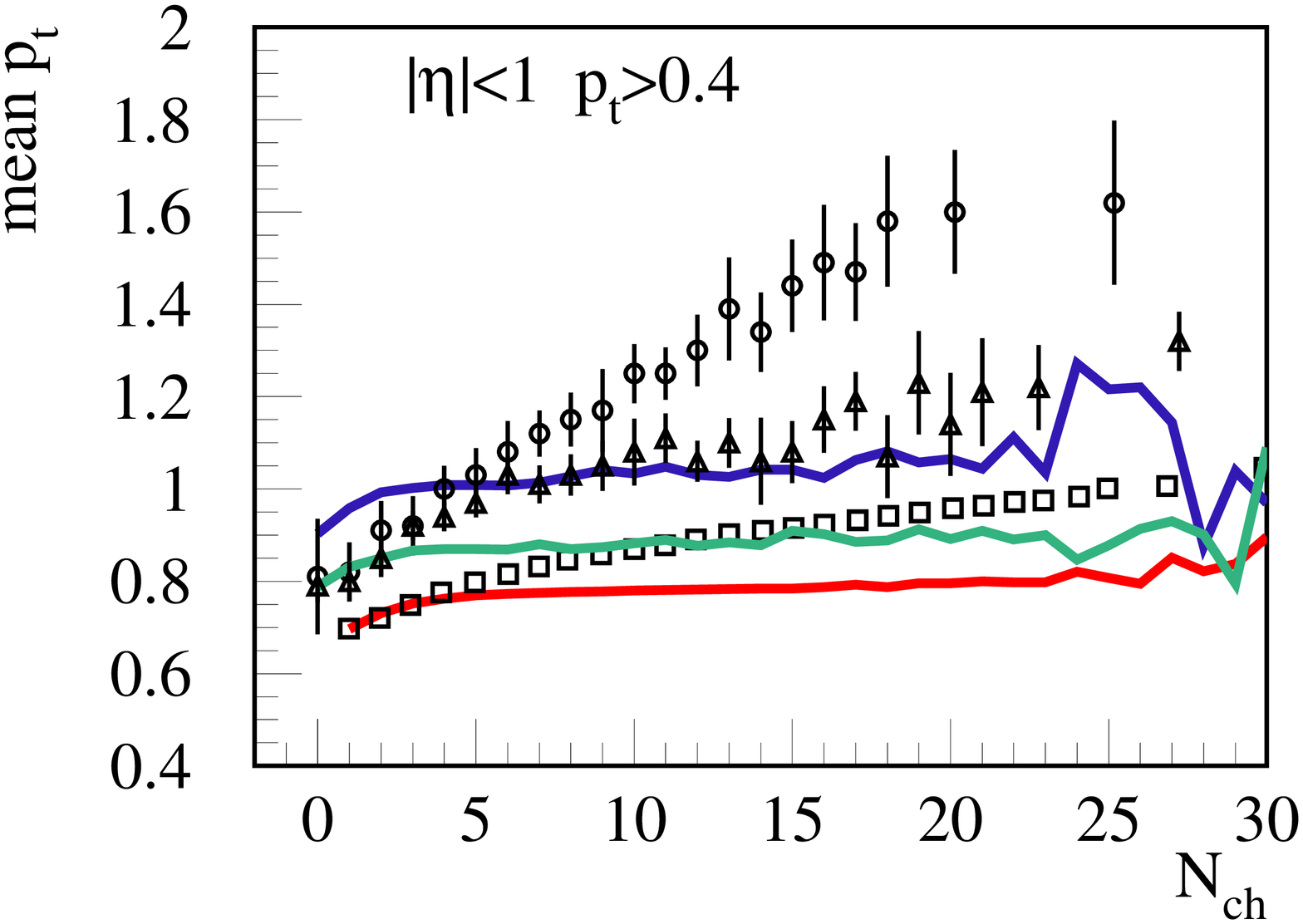}
\hfill
\includegraphics[width=0.38\textwidth,angle=-90]{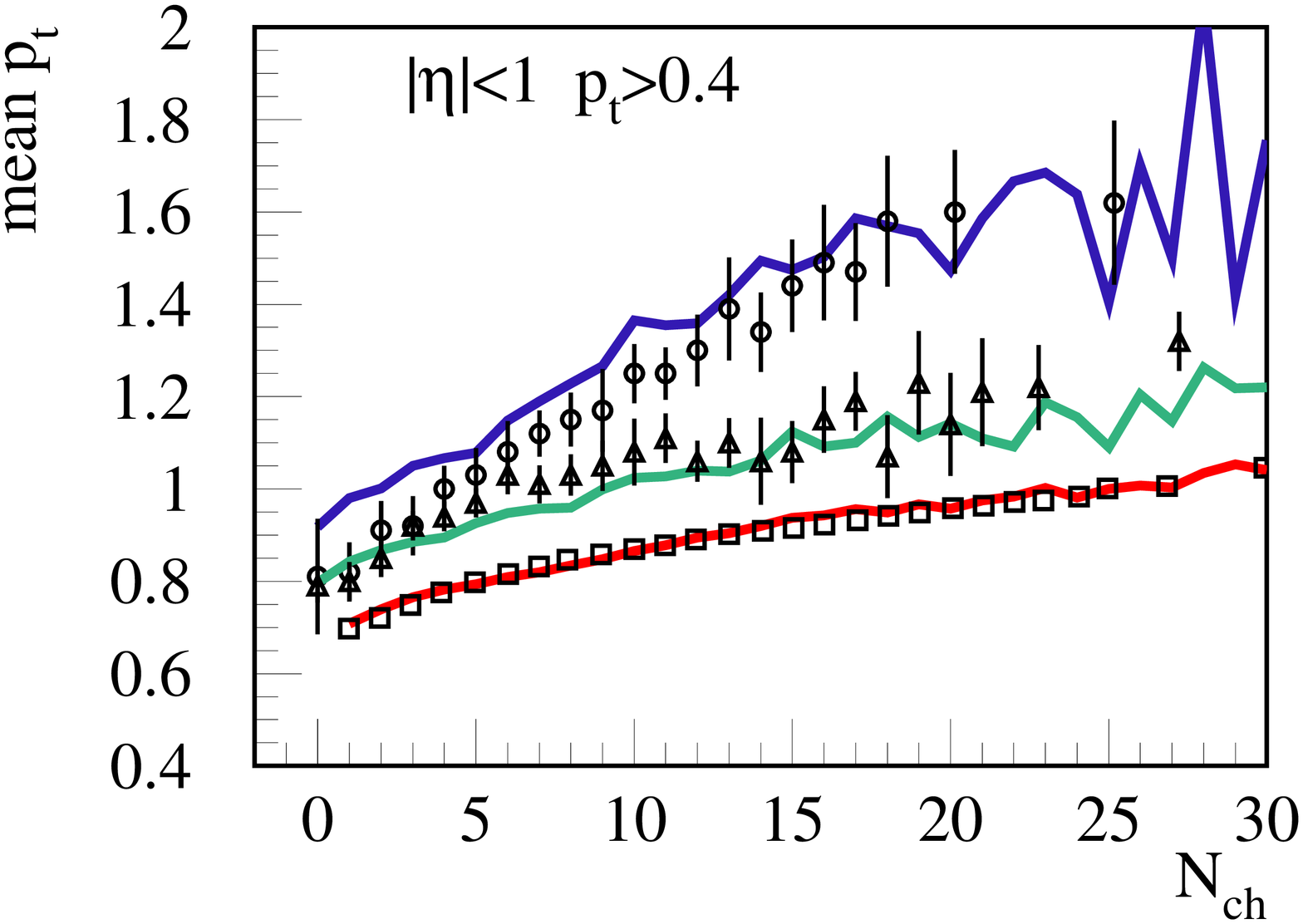}
\end{center}
\vspace*{-0.5cm}
\caption{Average transverse momentum $\langle p_t\rangle$ as a function of central multiplicity of
charged particles
for different particle types (from top to bottom : lambdas, kaons short and
charged particles). Points are data from CDF experiment~\protect\cite{Acosta:2005pk}.
Line are simulations with EPOS without hydro (left hand-side) and with hydro
(right hand-side).  \label{fig:tevatron}}
\end{figure}

As it was pointing out already in 2007 in~\cite{Abreu:2007kv}, a model which
describe properly these effects for heavy ion interactions will predict
the same mechanism already for (anti)proton-proton interactions when the
``centrality'' is large enough (high multiplicity in the central region).
And it can be compared to Tevatron data~\cite{Acosta:2005pk} where such effects 
are clearly visible
in the dependence of the average transverse momentum with central multiplicity
(including mass splitting) as shown fig.~\ref{fig:tevatron}.

After a short introduction on the EPOS model section~\ref{epos}, 
we will compare its results to the latest ATLAS data at 900 GeV 
and show section~\ref{ATLAS} that the same effect is already visible
event at this relatively ``low'' energy. In the section~\ref{hydro}, we will 
present how a correct calculation of 
collective effects can be done in the framework of the EPOS model.

\section{EPOS Model}\label{epos}

One may consider the simple parton model to be the basis of high energy 
hadron-hadron interaction models, which can be seen as an exchange of a 
``parton ladder'' between the two hadrons.

In additions to the parton ladder, hadronized using strings (flux-tube), 
there is another source of particle production: the two off-shell remnants.

EPOS~\cite{epos} is a consistent quantum mechanical multiple scattering approach
based on partons and strings, where cross sections
and the particle production are calculated consistently, taking into
account energy conservation in both cases.
Nuclear effects related to Cronin transverse
momentum broadening, parton saturation, and screening have been introduced
into EPOS. Furthermore, high density effects leading
to collective behavior in heavy ion collisions are also taken into
account. In next section preliminary results are shown using an effective 
treatment using a parameterized flow (not using LHC data). The full
hydrodynamic treatment described in section~\ref{hydro} applied to 
(anti)proton-proton will be shown in a paper in preparation.

\section{Comparison to LHC data}\label{ATLAS}
\begin{figure}[htbp]
\begin{centering}
\includegraphics[scale=0.22]{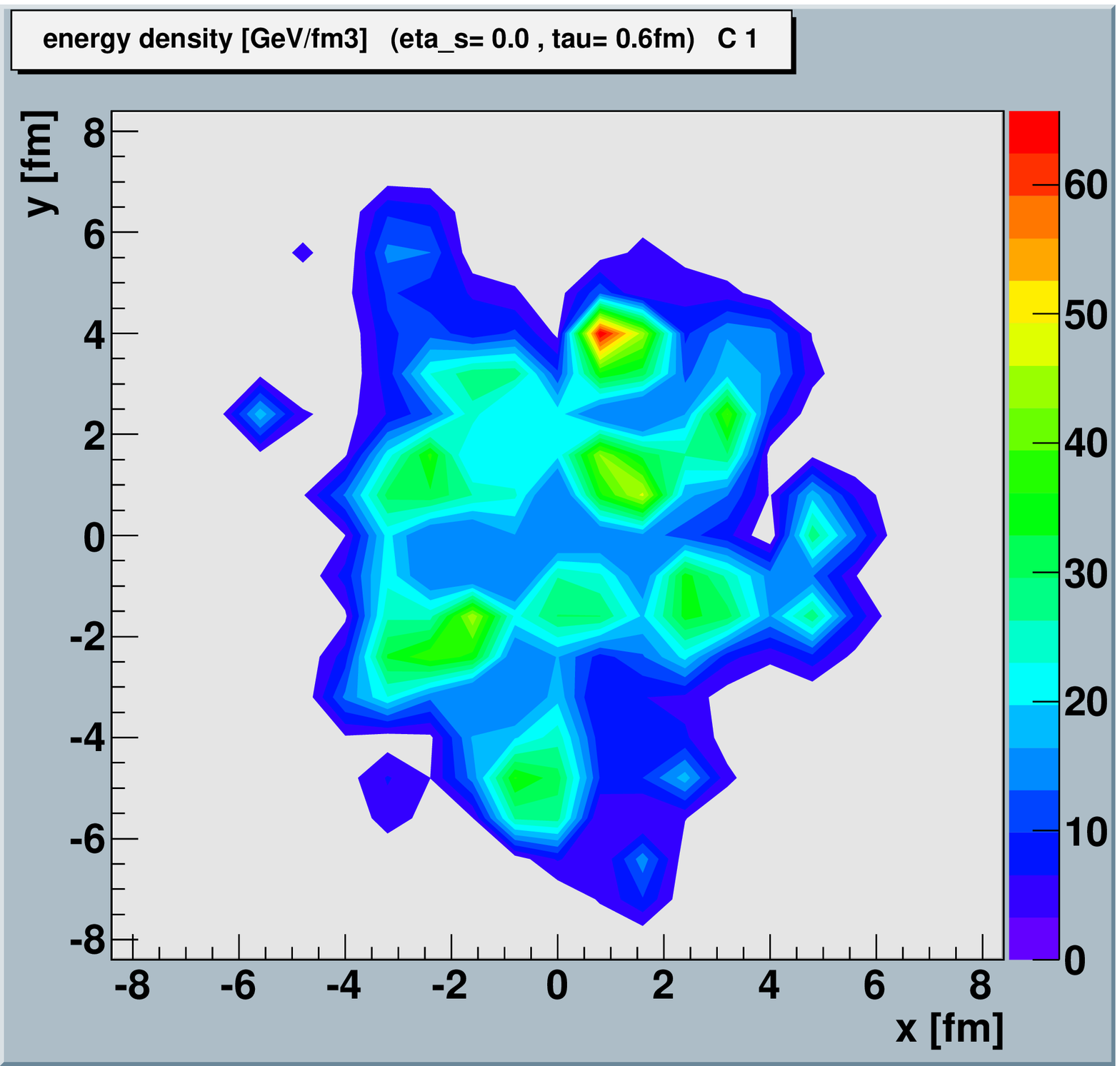}\includegraphics[scale=0.22]{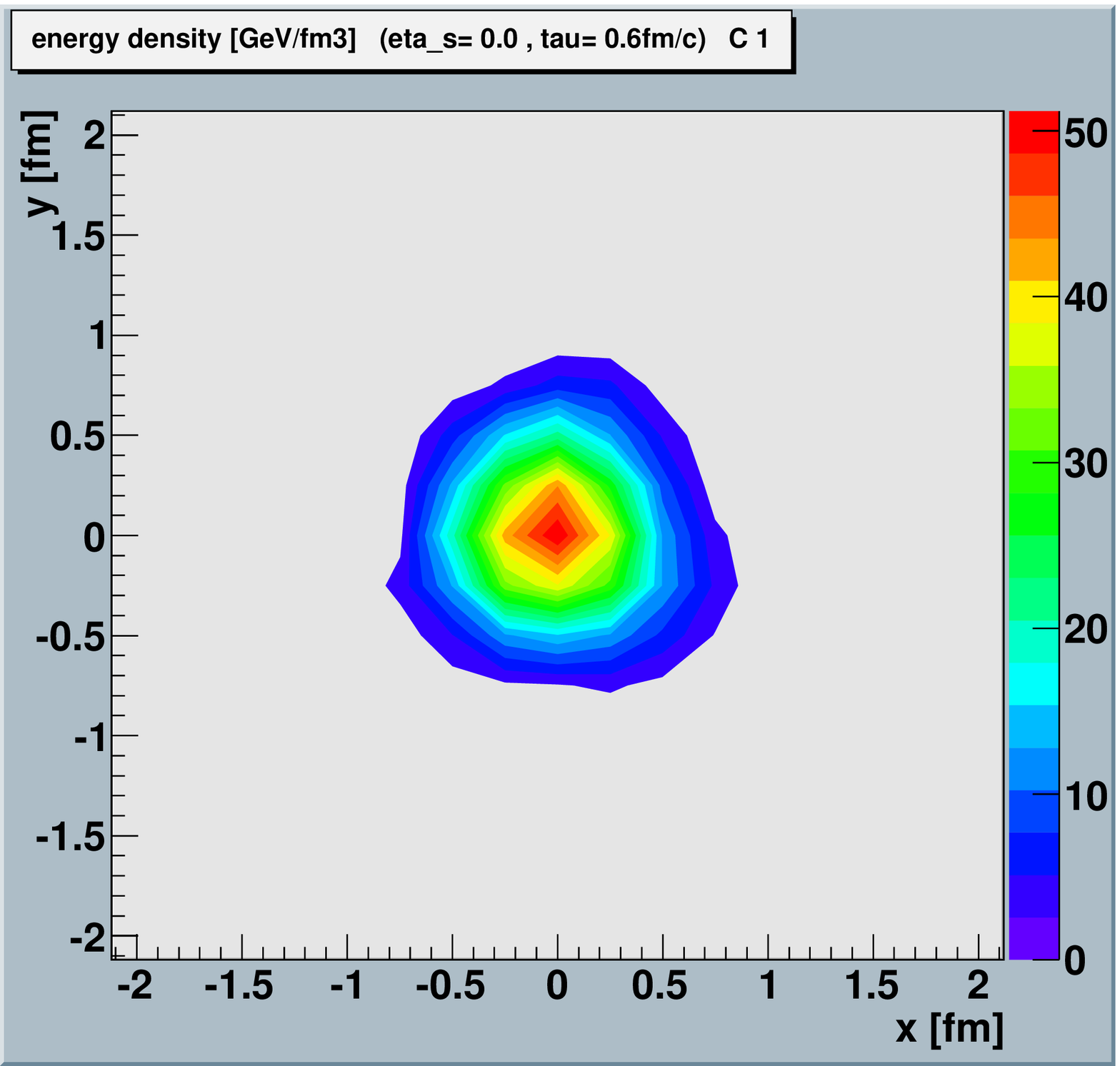}
\par\end{centering}

\caption{Energy density in central Au-Au (L) and pp (R) scattering at 200\,GeV 
resp. 7\'TeV.\label{fig:Energy-density-in}}
\end{figure}

Let us consider the energy density at an early time in a Au-Au scattering
at RHIC, as obtained from an EPOS simulation~\cite{epos}. 
In the left hand-side of fig.~\ref{fig:Energy-density-in}, 
we plot the energy density at the
 space-time rapidity $\eta_{s}=0$, as a function of the transverse
coordinates $x$ and $y$ . We observe a very bumpy structure concerning
the $x-y-$dependence.
There are in particular peaks in the $x-y-$plane which are sub-flux-tubes
which exhibit a long range structure in the longitudinal variable
$\eta_{s}$.  

In fig. \ref{fig:Energy-density-in} , we clearly identify several
sub-flux-tubes, with a typical width of the order of a fermi. This
is exactly the width we obtain if we compute the initial energy density
in proton scattering at the LHC (fig.~\ref{fig:Energy-density-in} 
right hand-side). This means, if a hydrodynamic treatment
is justified for Au-Au collisions at RHIC, it is equally justified
for pp scattering at the LHC, provided the energy densities are high
enough. This latter condition can easily be satisfied, since in proton-proton
scattering on has the possibility to trigger on high multiplicity
events, with ten or twenty times the multiplicity compared to an average
event.

If collective effects are possible in light system, how can it be
observed ? Besides correlations between particles, one of the striking 
consequence of
a collective hadronization is the creation of a radial velocity. In case
of heavy ion collisions, the asymmetry created by the impact parameter allows
the measurement of the $v_2$ parameter of the flow to quantify this effect.
In case of anti(proton)-proton scattering, it is experimentally difficult 
to define a ``collision plane'' and actually it was never done or even foreseen
 to measure $v_2$ in that case. But the radial flow increase the transverse 
momentum of the particles, so that this effect should change the mean 
transverse momentum $\langle p_t\rangle$. A measure of the energy density is 
given by the multiplicity of charged particles $N_{ch}$ event-by-event. 
As a consequence the variation of the $\langle p_t\rangle$ as a function 
of $N_{ch}$ has to be sensitive to the collective effects in a very light 
system.

\begin{figure}[htbp]
\begin{minipage}{0.47\textwidth}
\begin{center}
\includegraphics[width=0.98\textwidth]{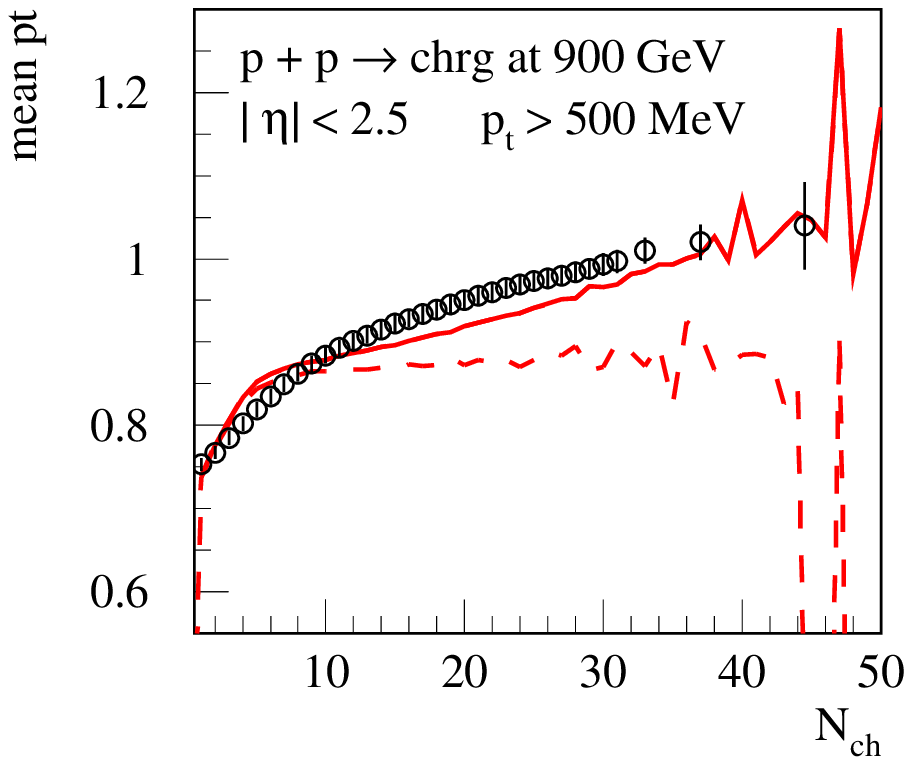}
\end{center}
\vspace*{-0.4cm}
\caption{$\langle p_t\rangle$ as a function of multiplicity of
charged particles
for proton-proton collisions at 900~GeV. ATLAS
data points~\protect\cite{Aad:2010rd} are compared to EPOS simulations with 
(full line) or without (dashed line) hydro evolution.\label{fig:apt}}
\end{minipage}
\hfill
\begin{minipage}{0.47\textwidth}
\begin{center}\includegraphics[width=0.98\textwidth]{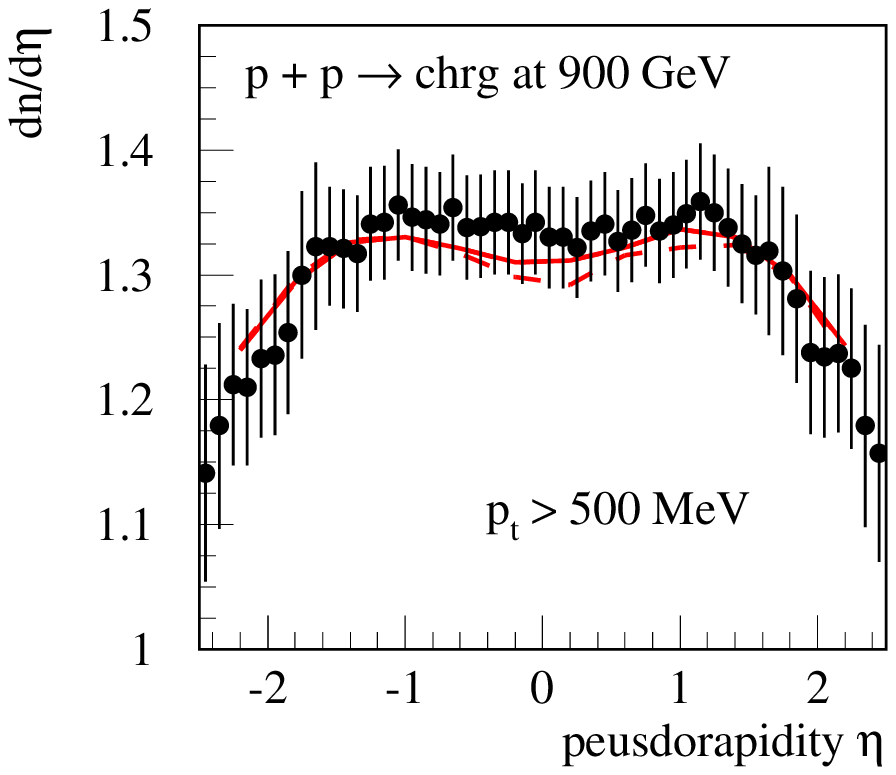} \end{center}
\vspace*{-0.4cm}
\caption{Pseudorapidity distribution of charged particles with p$_t>500$\,MeV
for proton-proton collisions at 900\,GeV. ATLAS
data points~\protect\cite{Aad:2010rd} are compared to EPOS simulations with 
(full line) or without (dashed line) hydro evolution.\label{fig:eta}}
\end{minipage}
\end{figure}

The effect shown in fig.~\ref{fig:tevatron} for Tevatron at 1.8\,TeV is 
actually confirmed at 900\,GeV by the ATLAS experiment~\cite{Aad:2010rd}. In
fig.~\ref{fig:apt} we can see the difference between simulation with 
(full line) or without (dashed line) hydrodynamic evolution of the high 
density region. This effect has only little influence on the total 
multiplicity has shown in fig.~\ref{fig:eta}.

\section{Hydrodynamics in EPOS}\label{hydro}

In future version of EPOS, we are going to employ a new tool for treating 
the hydrodynamic evolution,
based on the following features (see \cite{hydro} for details and tests
with AuAu data):

\begin{itemize}
\item initial conditions obtained from a flux tube approach (EPOS), compatible
with the string model used since many years for elementary collisions
(electron-positron, proton proton), and the color glass condensate
picture; 
\item consideration of the possibility to have a (moderate) initial collective
transverse flow;
\item event-by-event procedure, taking into the account the highly irregular
space structure of single events, being experimentally visible via
so-called ridge structures in two-particle correlations; 
\item core-corona separation, considering the fact that only a part of the
matter thermalizes;
\item use of an efficient code for solving the hydrodynamic equations in
3+1 dimensions, including the conservation of baryon number, strangeness,
and electric charge; 
\item employment of a realistic equation-of-state, compatible with lattice
gauge results -- with a cross-over transition from the hadronic to
the plasma phase; 
\item use of a complete hadron resonance table, making our calculations
compatible with the results from statistical models;
\item hadronic cascade procedure after hadronization from the thermal system
at an early stage.
\end{itemize}

\section{Summary}

EPOS is an interaction model constructed on a solid theoretical
basis. It has been tested very carefully against all existing hadronic
data. Designed to be compared to heavy ion collisions, the collective effects
are being implemented in a very sophisticated way. Based on a realistic 
event-by-event energy density with large fluctuations, a 3D hydrodynamical 
calculation is performed until chemical freeze-out followed by a hadron cascade
until thermal freeze-out. Applying the same scheme for (anti)proton-proton
scattering at high energy, the most inelastic collisions will actually satisfy
the conditions to create a small thermalized system which will hadronized 
statistically and with a non-negligible radial velocity. We showed that this 
effect is visible in the dependence of the $\langle p_t\rangle$ with the 
particle multiplicity of each event at Tevatron or LHC energies, especially
if we look at the dependence with the mass of the particles. Further 
detailed measurements at LHC (correlations) will allow us to actually test 
the space-time evolution of the energy density of hadronic interactions.

\section*{Acknowledgments}

This research has been carried out within the scope of the ERG (GDRE)
{}``Heavy ions at ultra-relativistic energies'', a European Research
Group comprising IN2P3/CNRS, Ecole des Mines de Nantes, Universit\'e
de Nantes, Warsaw University of Technology, JINR Dubna, ITEP Moscow,
and Bogolyubov Institute for Theoretical Physics NAS of Ukraine. Iu.
K. acknowledges partial support by the MESU of Ukraine, and Fundamental
Research State Fund of Ukraine, agreement No F33/461-2009. Iu.K. and
K.W. acknowledge partial support by the Ukrainian-French grant {}``DNIPRO\char`\"{},
an agreement with MESU of Ukraine No M/4-2009. T.P. and K.W. acknowledge
partial support by a PICS (CNRS) with KIT (Karlsruhe).

\end{document}